# FEATURES OF THE SECOND-ORDER VISUAL FILTERS SENSITIVE TO THE SPATIAL FREQUENCY MODULATIONS

**Miftakhova M.B., Babenko V.V.**
Southern Federal University
ms.miftakhova@mail.ru

*We investigated selectivity of $2^{nd}$ order visual filters sensitive to the spatial frequency (SF) modulations to orientation and to SF of modulation. We carried out psychophysical experiment using masking paradigm. It was found that $2^{nd}$ order filters are selective to SF of modulation (with bandwidth $\pm 1.5$ octaves), and do not show any selectivity to orientation of modulation. We suppose receptive fields of $2^{nd}$ order mechanisms have concentric form.*

## Introduction

The initial stage of processing visual information includes local linear filtering, which results in segregating of so-called primitives. Before selective attention turns on there is another stage where primary elements are united into "cognitive blocks". This synthesis is processed by the mechanisms of the second order [eg. 1, 2]. These mechanisms are sensitive to the spatial modulations of primary image features such as orientation, spatial frequency and contrast. The standard model describing processing of second-order visual stimuli is "filter - rectify - filter" model (FRF, "back-pocket model", Fig.1)

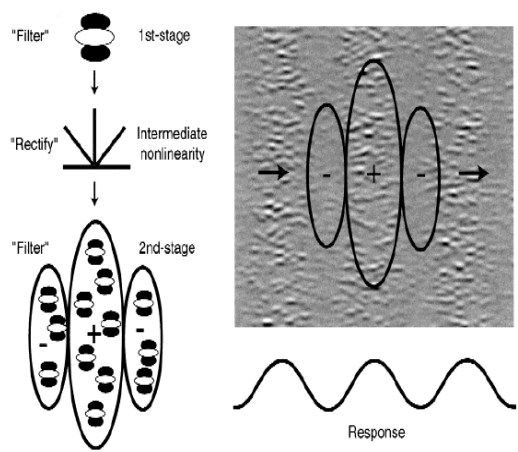

Fig.1 Second-order filter organization. First order filters are linear and have narrow tunings. Outputs of the first stage are inputs for non-linearity (full-wave rectification). Second-order filters unite outputs of the non-linearity stage and encode new, more complicated image features.

The aim of this work was to investigate tunings of second-order mechanisms sensitive to the spatial frequency modulations. Two experimental series were carried out to measure selectivity of the mechanisms to spatial frequency and orientation of modulation respectively.

## Methods

Our study was organized using masking paradigm, 2-alternative forced choice (2AFC) procedure, and the stair-case method.

Stimuli were textures composed of Gabor micropatterns (Fig. 2).

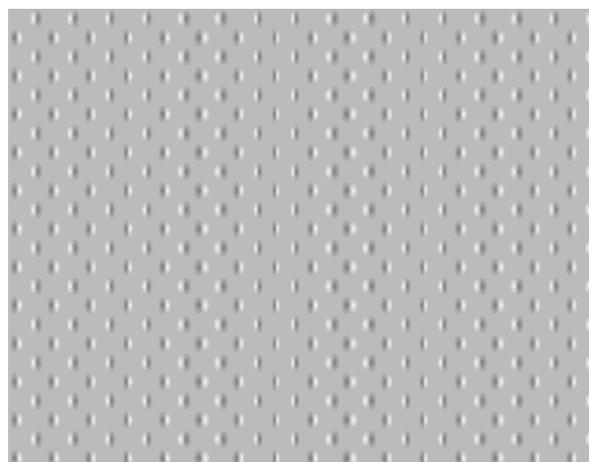

Fig.2 Example of stimulus used in the study. Texture composed of Gabor elements, sinusoidally modulated by spatial frequency.

Target stimulus was the texture sinusoidally modulated by spatial frequency.

In the experimental series with measuring tunings to orientation of modulation 5 masks were presented. Masking stimuli were textures with orientation range from 0 to 90 deg. with 22.5 deg. increment.

In the experiment with measuring tunings to spatial frequency of modulation we used 5 masks with frequencies from -3 to +3 octaves from the initial frequency, 1.5 octaves increment.

The subjects were placed at a distance of 130 cm from the display. Angle dimensions of the screen were 14x10.5 degrees. First test, then masking texture were performed in each of two time windows separated by an interval of 750 ms. Test duration was 250 ms, the duration of the mask was also 250 ms. One of these windows included modulated test texture, another included texture without any modulations.

Three subjects with normal or corrected to normal vision participated in the study.

For each observer more than 20 samples for each combination of stimulus-mask were registered.

## Results

*Selectivity to the spatial frequency of modulation*

Experiment with studying selectivity to the frequency of modulation showed significant differences in the amplitude thresholds for different types of masks (Fig. 3).

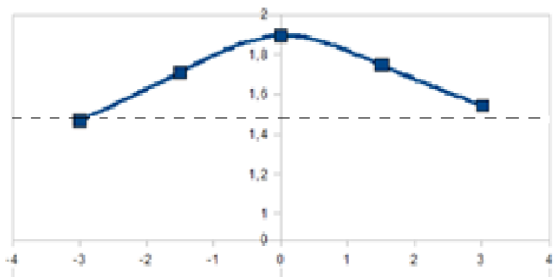

Fig.3 Summary data for three subjects. Abscissa shows spatial frequency of the mask (octaves relative to the test stimulus); the vertical axis shows the threshold amplitude in arbitrary units. Broken line shows amplitude threshold for control mask (without modulations).

The maximum masking effect was observed when test and mask textures where identical, minimum thresholds were found at frequencies of $\pm 3$ octaves. These results indicate the presence of selectivity of the second-order mechanisms sensitive to spatial frequency modulations. The bandwidth at half amplitude is equal to $\pm$ 1.5 octaves.

.
*Selectivity to orientation of modulation*

Studying the selectivity of second-order mechanisms to orientation of modulation showed no significant differences in threshold amplitudes for different types of masks (Fig. 4).

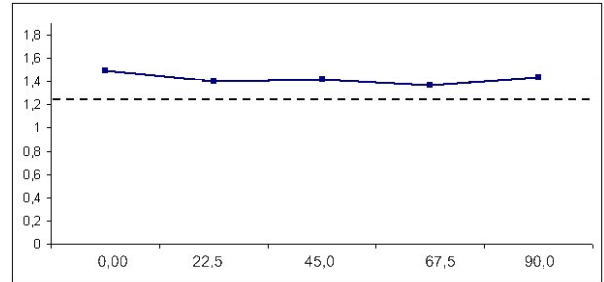

Fig. 4 Threshold amplitude changing for masks with different modulation orientations. Abscissa shows the axis tilt of modulation in the masking stimulus, ordinate shows modulation threshold in arbitrary units. Broken line shows amplitude threshold for control mask (without modulations).

The graph shows no significant change in threshold amplitude for different orientations of modulation. Thus we didn't find the selectivity of second-order mechanisms to the orientation of modulation.

## Conclusion

We investigated the selectivity of the second-order visual filters sensitive to spatial frequency modulations to the frequency and orientation of the modulation.

In recent studies the selectivity of second-order filters sensitive to the contrast modulations was shown. Measured bandwidth was equal to $\pm 1$ octave [4], what is comparable with the bandwidth for the first-order mechanisms [5]. These filters are selective to orientation and phase of modulation. According to this fact we suppose that second-order filters sensitive to the contrast modulations have receptive fields (RF) with elongated form and inhibitory subfields.

Filters sensitive to the spatial frequency modulations cannot have elongated form of RFs because of absence of orientation selectivity. More over filters under study have broader bandwidth. These distinctions can be explained by another spatial organization of receptive fields of such mechanisms.

The lack of orientation selectivity of filters sensitive to spatial frequency modulations and broad tuning to spatial frequency of these filters suggest a possibility of a concentric organization of their RFs.